\documentclass[%
 reprint,
 superscriptaddress,
 amsmath,amssymb,
 aps,
 pre,
]{revtex4-2}

\usepackage{graphicx}
\usepackage{dcolumn}
\usepackage{bm}
\usepackage{hyperref}
\usepackage{color}
\usepackage{braket}

\renewcommand{\selectlanguage}[1]{}

\makeatother

\begin{document}

\title{$\mathcal R$-Ising: Effective resistance in random magnetic nanowires networks}

\author{Frank Barrows}
\affiliation{Center for Nonlinear Studies, Los Alamos National Laboratory, Bikini Atoll Road, Los Alamos 87544, NM (USA)}
\affiliation{Theoretical Division (T4), Los Alamos National Laboratory, Bikini Atoll Road, Los Alamos 87544, NM (USA)}

\author{Ezio Iacocca}
\affiliation{Center for Magnetism and Magnetic Nanostructures, University of Colorado Colorado Springs, Colorado Springs 80918, CO (USA)}

\author{Francesco Caravelli}
\email{caravelli@lanl.gov}
\affiliation{Theoretical Division (T4), Los Alamos National Laboratory, Bikini Atoll Road, Los Alamos 87544, NM (USA)}

\date{\today}

\begin{abstract}
Random assemblies of magnetic nanowires represent a unique class of materials with promising applications in spintronics and information storage. These assemblies exhibit complex behavior 
due to the combination of magnetic dipolar interactions between the nanowires and electronic transport properties governed by tunneling barriers at magnetic tunnel junctions (MTJs). The intricate interplay of these phenomena makes the study of magnetic nanowire networks a rich area of research. In this study, we develop a theoretical framework to analyze the resistive behavior of random magnetic nanowire networks. By employing a combination of graph theoretical approaches and mean-field theory, we derive an effective resistance model that encapsulates the contributions of magnetic interactions between the nanowires. Our findings show the importance of considering both the magnetic and electrical properties of nanowire networks in the design and optimization of amorphous resistive devices.
\end{abstract}

\maketitle

\section{Introduction}
Random assemblies of magnetic nanowires represent a unique class of materials with promising applications in various fields, including spintronics and information storage \cite{Fert1999,Soumare2008,Mukhtar2020,Stao2018,Piraux2020}.
These assemblies exhibit complex interactions, primarily due to the combination of magnetic dipolar interactions between the nanowires.
When two nanowires intersect, they form a junction separated by a thin insulating layer~\cite{Brajuskovic_APLMaterials_2022}, acting as a magnetic tunnel junction (MTJ). The electronic transport properties of the MTJ are governed by tunneling barriers 

Specifically, in the context of neuromorphic materials, silver-based nanowires have been employed in the study of atomic switch networks \cite{Scharnhorst2018,DiazAlvarez2019,Aguilera2020,Lilak2021}. Atomic switch networks (ASNs) are a class of nanostructured materials characterized by their dynamic and reconfigurable electrical properties. These networks are typically based on silver (Ag) due to their excellent electrical conductivity and the ability to form conductive filaments at the nanoscale. The chemical synthesis of ASNs involves several steps, including the preparation of a suitable substrate, the deposition of silver nanoparticles, and the formation of atomic switches. 

The resulting ASN exhibits unique features such as non-volatile memory, reconfigurable conductivity, and potential applications in neuromorphic computing\cite{Demis_Nanotechnology_2015,Avizienes_PLOS_2012,Bhattacharya_NanoLett_2022}. Currently, a variety of different machine learning tasks can be performed using these types of nanowires, ranging from classification to time series prediction \cite{Zhu2021,Zhu2023}, and in particular emergent intelligence  \cite{Tsuchiya2022}. In analogy with the current literature on artificial connectomes \cite{Milano2022a,Milano2022b}, the complex behavior of these materials can be used to improve the learning capability of these devices \cite{Dunham2021,Caravelli2023,Baccetti2024,Caravelli2021}. One of the drawbacks of silver based nanowires is that silver is not a magnetic material. Thus, the resistive behavior of the device is
purely emergent and hardly controllable.

On the other hand, magnetic nanowires, and particularly those made from cobalt (Co) or permalloy (Ni$_{80}$Fe$_{20}$), have gained interest for their potential in spintronics and magnetic storage devices \cite{Hirohata2015}. These materials exhibit strong magnetic properties and can be fabricated using various methods, including electrodeposition, template-assisted growth, and focused electron beam-induced deposition (FEBID)\cite{Fernandez-Pacheco_SciRep_2013,Pablo-Navarro_IOP_2017,Ruiz-Gomez2022-ha}. The electrodeposition method involves using a porous template, such as anodized aluminum oxide (AAO), which contains an array of nanoscale pores\cite{Brajuskovic_APLMaterials_2022,Carignan_IEEE_2011}. The template is immersed in an electrolyte solution containing metal ions, and an electric current is applied to drive the deposition of metal into the pores. This results in the formation of nanowires with high aspect ratios and controlled dimensions. 

After deposition, the template is dissolved, leaving behind an array of free-standing nanowires.
For magnetic nanowires made from permalloy, the electrodeposition process involves a solution containing both nickel and iron ions. By carefully controlling the deposition parameters, it is possible to achieve the desired stoichiometry and magnetic properties. Cobalt nanowires can be fabricated similarly, using a Cobalt salt solution as the electrolyte\cite{Hoshino_ElectrochemComm_2005,Balela_ElectrochemSoc_2011}.

These magnetic nanowires exhibit interesting properties such as high coercivity, anisotropic magnetoresistance, and the potential for use in high-density magnetic storage devices and spintronic applications. The ability to precisely control the dimensions and composition of these nanowires makes them attractive for research and development in advanced magnetic materials. Similarly, the density of nanowires is a controllable parameter. 

In magnetic amorphous and ordered assemblies \cite{Caravelli2022a,Caravelli2022b}, the nanowires are often fully magnetized, with their magnetization pointing in either direction along the wire. As mentioned earlier, the resistance of these junctions depends on 
the relative orientation of the magnetization of the intersecting nanowires, with the resistance being different for ferromagnetic and antiferromagnetic alignments \cite{Caravelli2018,Sun2000}. This dependence introduces a level of complexity in understanding the overall electrical behavior of the network.

The energy landscape of these networks can be described by a Hamiltonian that takes into account both the magnetic interactions and the resistive states of the junctions. The magnetic energy is primarily dipolar in nature, and domain walls can nucleate at finite temperatures, affecting the magnetization states of the nanowires\cite{Brajuskovic_APLMaterials_2022,Sanchez_PRB_2017,Maurer_JAP_2011}. These domain walls play a critical role in determining the effective resistance of the network, as they can propagate through the nanowires and flip their magnetization.
\begin{figure}
    \centering
    \includegraphics[width=0.99\linewidth]{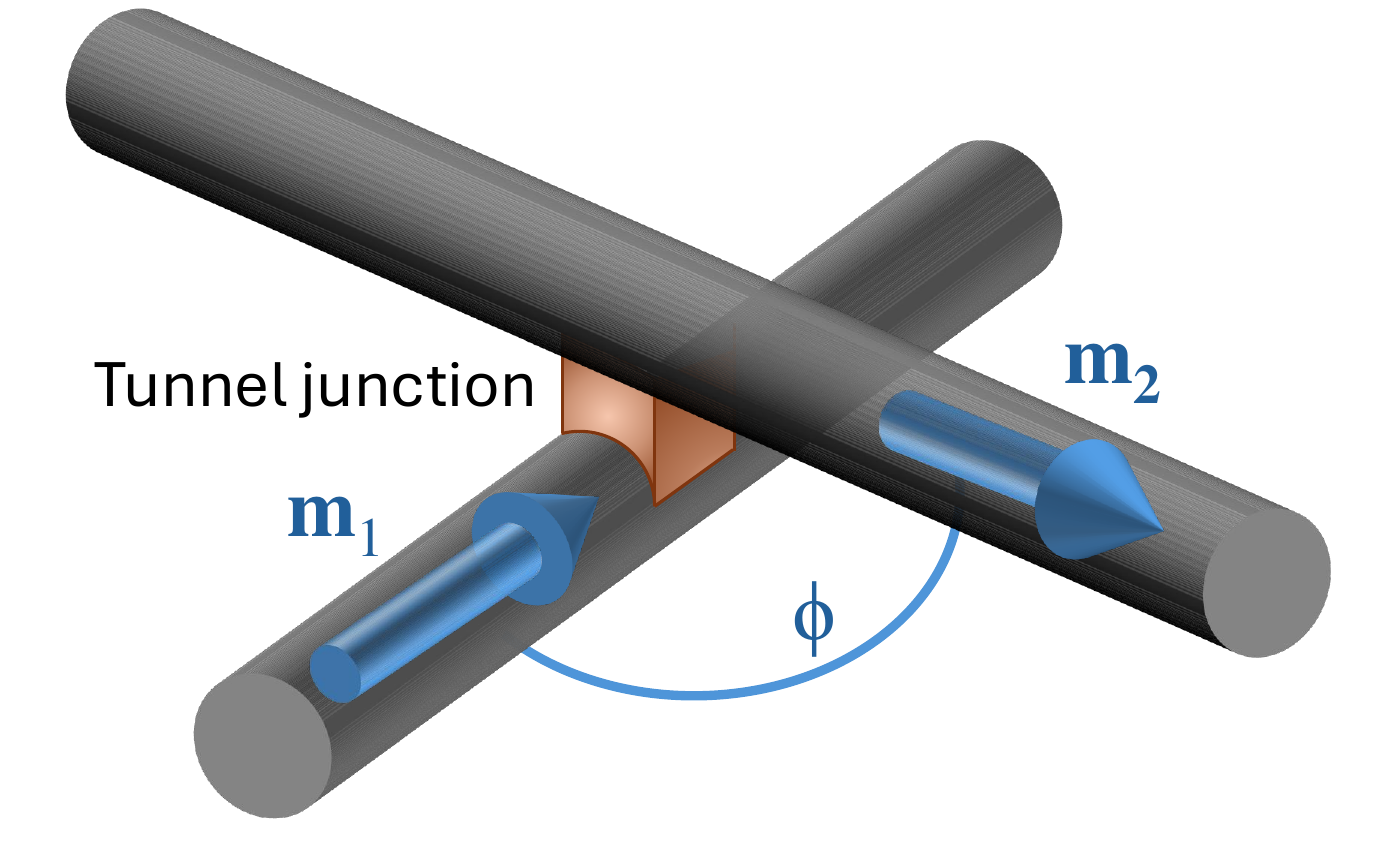}

    \caption{The magnetic tunnel junction (MTJ) at the avoided crossing of two nanowires. The blue arrows represent the magnetization vector of each nanowire, assumed to be uniform. The angle between the nanowires is $\phi$. }
    \label{fig:mtj}
\end{figure}
Furthermore, the resistance of a junction is not static but depends on the magnetic states of the nanowires forming a junction. 
This can be mathematically expressed by considering the contributions of ferromagnetic and antiferromagnetic states, leading to a resistance formula that encapsulates the effects of the magnetization states. For practical purposes, it is often more convenient to work with the conductance rather than resistance, enabling a more straightforward analysis of the network properties.

The effective resistance of the entire network is a critical parameter that influences its performance in applications. This resistance can be calculated using techniques such as the Moore-Penrose pseudoinverse of the Laplacian matrix of the network graph, which represents the nanowires as nodes and the junctions as resistive edges. The resulting expressions provide insights into how the network responds to external stimuli and how the magnetic interactions influence its electrical characteristics.

In this paper, we explore the physical properties of random magnetic nanowire networks with a focus on understanding the role of magnetic tunnel junctions. We analyze the network's resistive behavior by considering the interplay between magnetic and electrical properties, providing a comprehensive framework for predicting the performance of these novel materials in practical applications.

\section{The model}
We consider the following model to determine the resistive state of a magnetic nanowires disordered system. We assume that the magnetic nanowires are in a fully magnetized state with the magnetization along the wire's length. 
We assume that there is a certain angle {$\phi_{ij}$} between two nanowires $i,j$, with a thin insulating interlayer. As a result, the avoided crossing between the nanowires acts as a magnetic tunnel junction. The resistance of the junction depends on the orientation of the magnetization $\vec M_{i}$ and $\vec M_j$, with  $\vec M_j=M \vec{s}_j $. It is very well known that at the crossing between the wires, domain walls can nucleate at a finite temperature and propagate through the wire, effectively flipping the magnetization \cite{Caravelli2022a}. Thus, given a certain orientation of the nanowires, it is our assumption going forward that the magnetization is only oriented along the nanowires, thus taking values $\vec M_j$ and $-\vec M_j$. 

As we see later, we can thus think of the island's magnetization as being proportional to an Ising spin, while being careful in incorporating the angles into the interaction constants. As a result, the energy of the magnetic network can be written as $E=\sum_{ij} J_{ij} {s}_i {s}_j$ at a first level of approximation. The interaction $J_{ij}$ incorporates the angle $\phi_{ij}$ between the nanowires, i.e., $J_{ij}$ depends on $ \cos(\phi_{ij})$, assuming that the magnetic energy is dipolar in nature.  If we for instance assume that islands are on top of each other, then $\hat r_{ij}$, the vector in the direction of their geometrical centers, will be perpendicular to their direction $\vec s_i$ (assumed to be in plane). Thus, since $\hat s_i\cdot \hat r_{ij}\approx 0$ and $\hat s_j\cdot \hat r_{ij}\approx 0$, we have $J_{ij}\propto -J \cos(\phi_{ij})$. At finite temperatures, the nucleation of domain walls will be associated with the magnetic energy above. In addition, it has been experimentally observed that in etched thin films, currents strongly affect the heating, changing the resistivity of the devices \cite{Fonseca2022}.

Another interesting observation is that the resistance of the junction, assuming that the angle is fixed, is given by a function of the state of the magnetization of the nanowires, e.g., we have $R_{ij}=R(s_i,s_j)$. 

Magnetic tunnel junctions (MTJs) are fundamental components in spintronics, used in applications such as magnetic random-access memory (MRAM) and read heads for hard disk drives \cite{Miyazaki1995}. An MTJ consists of two ferromagnetic layers separated by a thin insulating barrier \cite{Fert1999}. The electrical resistance of the junction depends on the relative orientation of the magnetization in the two ferromagnetic layers due to spin-dependent tunneling.

\subsection{Model of Magnetic Tunnel Junctions}

Consider an MTJ with the top and bottom ferromagnetic layers represented by Ising spins $s_1$ and $s_2$, where $s_1, s_2 = \pm 1$. In the nanowire networks the Ising spin orientation corresponds to a magnetization vector confined to a two-dimensional plane. With uniform nanowires we can work with normalized spin vectors, $\vec{s}$. The relative angle between the magnetizations of the layers can be expressed as $\phi$ \cite{Caravelli2018}.

The conductance $G(\phi)$ of the MTJ is typically modeled as being higher when the magnetizations are parallel ($\phi = 0$) and lower when they are antiparallel ($\phi = \pi$) \cite{Sun2000}. This is because the density of states for electrons with different spin orientations at the Fermi level influences the tunneling probability \cite{Mathon2001}. For simplicity, we can assume that the conductance $G(\phi)$ takes binary values:
\begin{equation}
G(\phi) = \begin{cases}
G_f & \text{if } \phi=0 \text{ (parallel)}, \\
G_p & \text{if } \phi=\pi \text{ (antiparallel)},
\end{cases}
\end{equation}
where $G_f$ and $G_p$ are the conductances in the parallel and antiparallel configurations, respectively. 
Typical values for the resistance of an MTJ are around $\sim 10 \Omega$, with a change in resistance up to $50\%$ between the two logical values. These values must be generalized for the case in which the angle is not simply $0$ or $\pi$.

The angle $\phi_{ij}$ between the spins can be related to the Ising spin orientations as follows:
\begin{equation}
\vec{s}_i \cdot \vec{s}_j =  s_is_j\cos(\phi_{ij}),
\end{equation}
 with $\phi_{ij}$ the geometric angle between the nanowires in the network. 

Substituting $\cos(\phi)$ into the conductance equation, we obtain:
\begin{equation}
G(s_1, s_2) = \begin{cases}
G_f & \text{if } \vec{s}_1  \cdot \vec{s}_2 = 1, \\
G_p & \text{if } \vec{s}_1  \cdot \vec{s}_2  = -1.
\end{cases}
\end{equation}

The resistance $R(s_1, s_2)$ is the reciprocal of the conductance:
\begin{equation}
R(s_1, s_2) = \frac{1}{G(s_1, s_2)}.
\end{equation}
 This explains why $G_f$ and $G_p$ are influenced by the angle $\phi$, highlighting the angle's role in determining the conductance. 

In magnetic nanowires, the magnetization is typically aligned along the long axis of the nanowire. Therefore, the relative angle between the magnetizations of two nanowires is determined by their geometrical arrangement. These angles are fixed by the physical structure of the nanowires.  However, the magnetization of the nanowires can flip due to magnetic interactions, such as dipolar interactions. When the magnetizations of the nanowires flip, the relative angle between them changes, thus affecting the conductance of the MTJ. 

Since we assume that $R_a,R_f\gg R_n$, where $R_n$ is the resistance of each nanowire, we can convert the nanowire network to a network in which the resistances are the junctions and the nodes are the wires. This framework is valid when the magnetic nanowires are in a fully magnetized state, the presence of domain walls would require adopting a clique representation of each wire.

Under these conditions, it is more convenient to work with the conductances. We can write the conductance for each junction in terms of $G_f^{(ij)}$ and $G_a^{(ij)}$, corresponding to a ferromagnetic state or anti-ferromagnetic, as $G_{ij}=G_{f}(s_i,s_j) G^{(ij)}_f+P_{a}(s_i,s_j)G^{(ij)}_a$, where $P_f/P_a$ are projectors on those states, e.g. $P_f=\frac{(1-s_i)(1-s_j)}{4}+\frac{(1+s_i)(1+s_j)}{4}$, while $P_a=\frac{(1+s_i)(1-s_j)}{4}+\frac{(1-s_i)(1+s_j)}{4}$. We can then write 
\begin{equation}
    (G)_{ij}=g(i,j)=\frac{G^{(ij)}_p+G^{(ij)}_f}{2}+\frac{G^{(ij)}_f-G^{(ij)}_a}{2} s_i s_j
\end{equation}
where the $G^{(ij)}_{a/f}$ depend on $\cos(\phi_{ij})$.  Let us now discuss how to incorporate these resistances into the nanowire network. Below, we will use the notation in which pairs of nodes $\{(i,j)\}=\{\gamma\}$, and $\gamma(1)$ and $\gamma(2)$ as the first and second node.

\begin{figure}
    \centering
    \includegraphics[scale=0.4]{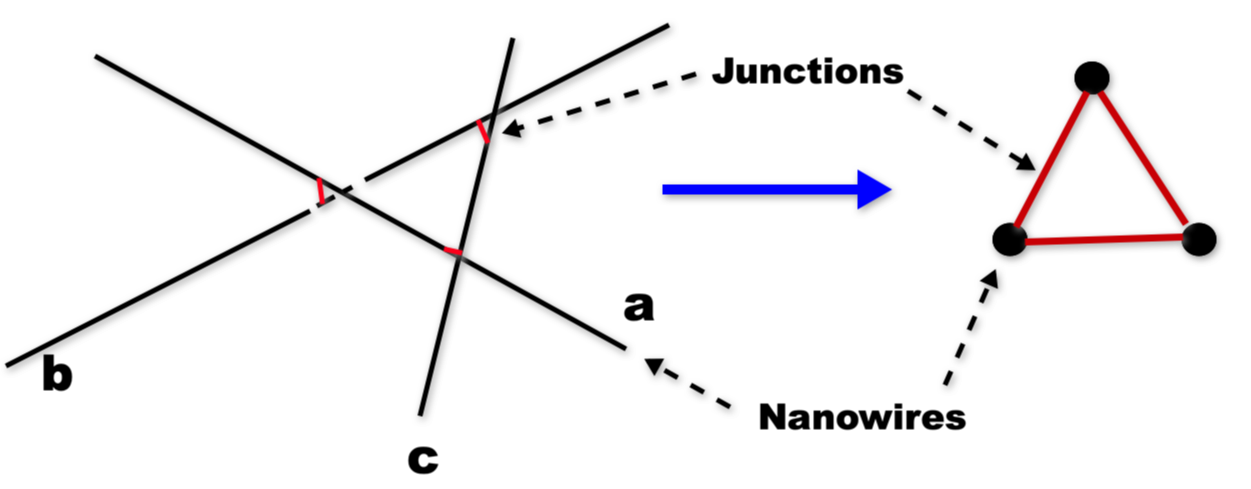}
    \caption{Mapping the nanowire network to the equivalent resistive graph. The avoided crossing becomes the edges of the graph (resistances), while the nanowires become the nodes. The same technique was used in \cite{Caravelli2023} in the case of silver nanowires.}
    \label{fig:junctmap}
\end{figure}

 \section{Effective resistance}
 \subsection{Graph theoretical approach to effective resistance}
 We now ask what is the effective resistance for this network. To be more specific, we consider a network of nanowires where each overlap 
 acts as a junction with higher resistance compared to the nanowires themselves. The nanowires are represented as nodes, and the junctions are represented as resistive edges in a graph, shown schematically in Figure \ref{fig:junctmap}. Let $V_i$ represent the potential at node $i$. Let $R_{ij}$ be the resistance of the junction between nanowires $i$ and $j$.
The current $I_{ij}$ through the junction is given by Ohm's law $I_{ij} = \frac{V_i - V_j}{R_{ij}}$. By Kirchhoff's current law (KCL), the sum of currents at node $i$ is zero 
    $\sum_{j} I_{ij} = 0$.
Formally,  we represent the graph $\mathcal{G} = (\mathcal{V}, E)$, where $\mathcal{V}$ is the set of nodes (wires) and $E$ is the set of edges (junctions). 
The adjacency matrix $A$ of the graph $\mathcal{G}$ has elements $a_{ij}$:
    \begin{equation}
    a_{ij} = 
    \begin{cases}
    1 & \text{if there is a junction between } i \text{ and } j \\
    0 & \text{otherwise}.
    \end{cases}
    \end{equation}
The conductance matrix $G$ with elements $G_{ij} = \frac{1}{R_{ij}}$ if $i$ and $j$ are connected, and $G_{ij} = 0$ otherwise. We assign an arbitrary orientation to each edge in the graph; the node-edge incidence matrix $B$ has elements:
    \begin{equation}
    B_{ik} = 
    \begin{cases}
    1 & \text{if edge } k \text{ points away from } i  \\
    -1 & \text{if edge } k \text{ points toward to } i  \\
    0 & \text{otherwise}
    \end{cases}
    \end{equation}

Given a network with the resistance matrix $R$ and the incidence matrix $B$, the effective resistance between nodes $i$ and $j$ can be computed as follows:
The resistance matrix $R$ is a diagonal matrix where each diagonal element $R_k$ represents the resistance of the $k$-th junction:
\begin{equation}
R = \text{diag}(R_1, R_2, \ldots, R_m) \rightarrow G=\text{diag}(G_1, G_2, \ldots, G_m)
\end{equation}
We define the Laplacian matrix $L$ given by:
\begin{equation}
L = B R^{-1} B^T=B G B^T .
\end{equation}
We can write the elements of this matrix as 
\begin{equation}
    L_{ij} = \begin{cases}
\sum_{k \,:\, \text{edge } k \text{ incident to } i} g_k & \text{if } i = j \\
-\sum_{k \,:\, \text{edge } k \text{ connecting } i \text{ and } j} g_k & \text{if } i \neq j
\end{cases}
\end{equation}
with $g_k$ the conductivity of the edge indexed $k$. Then the effective resistance $R_{ij}^{\text{eff}}$ between nodes $i$ and $j$, we use the Moore-Penrose pseudoinverse $L^+$ of the Laplacian matrix $L$:
\begin{equation}
R_{ij}^{\text{eff}} = (e_i - e_j)^T L^+ (e_i - e_j)
\end{equation}
where $e_i$ and $e_j$ are the standard basis vectors corresponding to nodes $i$ and $j$ in $\mathcal{G}$, respectively. 

Alternatively, we can use the elements of $L_{(1)}^{-1}$ to write:
\begin{equation}
R_{ij}^{\text{eff}} = L_{ii}^+ + L_{jj}^+ - 2L_{ij}^+
\end{equation}
where $L_{ii}^+$, $L_{jj}^+$, and $L_{ij}^+$ are the elements of $L^+$ corresponding to nodes $i$ and $j$. 
The mean effective resistance $\mathcal R_{\text{eff}}$ of the graph is the average of the effective resistances over all pairs of nodes:
\begin{equation}
\mathcal R = \frac{2}{n(n-1)} \sum_{1 \leq i < j \leq n} R_{\text{eff}}(i, j).
\end{equation}
It is known that the graph pseudo-inverse can be made invertible by considering a modified Laplacian. Below, we will use the following formula
\begin{equation}
    A^+=\lim_{\delta \rightarrow 0}(A^t A+\delta I)^{-1}A^t .
\end{equation}
Thus, we can obtain the Laplacian as a correlator. First, treating $A$ as the Laplacian, since $A=B G B^t$ is symmetric, we have $A^tA=A$. For symmetric matrices, it can be proven that the formula above reduces to a simpler version in this case, which is 
\begin{equation}
    A^+=\lim_{\delta \rightarrow 0}( A+\delta I)^{-1}.
\end{equation}
Then, 
we have
\begin{equation}
    A_{ij}=(B G B^t)_{ij}=\sum_{\gamma} B_{i\gamma}B_{\gamma j} G_\gamma
\end{equation}
where $\gamma$ is associated with a particular edge.
We will now discuss the averaging of the effective resistance
\begin{eqnarray}
    \langle L^+\rangle_\beta = \sum_{\{s_i\}} L^+(s_i) \frac{e^{-\beta H(s_i)}}{Z}
\end{eqnarray}
over spin configurations $\{s_i\}$. Calculating this quantity requires the use of the integral formula
\begin{equation}
    (L^+)_{ij}=\lim_{\delta\rightarrow 0}(B G B^t +\delta I)^{-1}_{ij}.
\end{equation}
which is the equation we use in the following section.

\subsection{Mean-field theory}
\begin{figure}
\includegraphics[scale=.2]{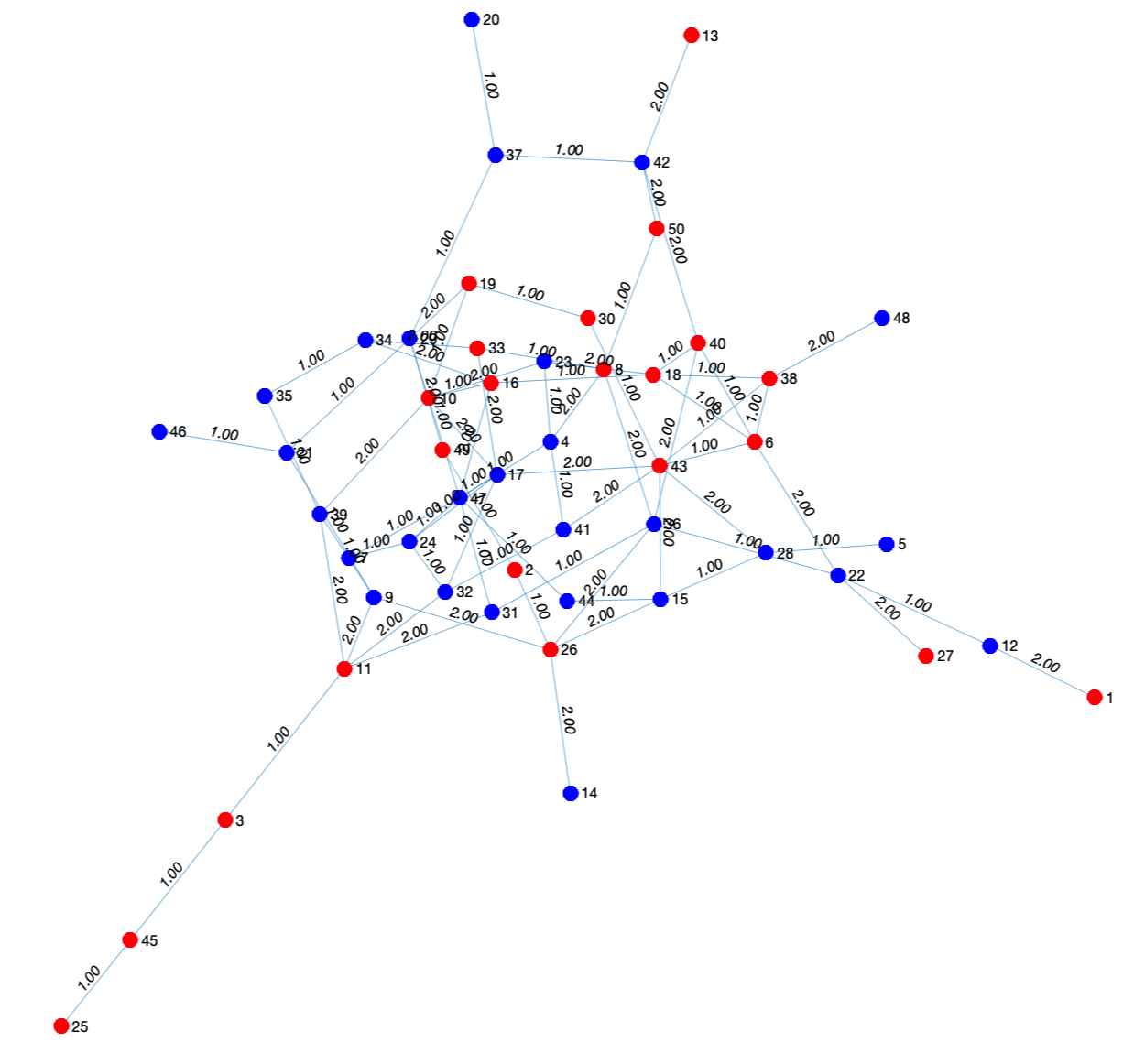}\\
\caption{A spin configuration of a random graph in the class $\mathcal G(n,m)$ with $n=50,m=80$.
 In this example, the resistance for opposite signs is $R_{+-}=2$, while for identical signs it is $R_{++}=1$. Red and blue nodes indicate the spin orientation, nodes are indexed and $R_{++}$ and $R_{+-}$ vales are given along the graph edges.}
\label{fig:RandonNetwork}
\end{figure}

It is challenging to determine the network properties for arbitrary networks of magnetic nanowires. Such an arbitrary network is shown in Figure \ref{fig:RandonNetwork}.
Clearly, the magnetic configuration of the nanowire network affects the resistivity of the network. If we want to estimate the average resistivity as a function of the temperature, we need to average the effective resistance. We will now use an annealed approximation,
using the formula:
\begin{eqnarray}
    M_{ij}^{-1}=\langle x_{i} x_{j}\rangle=\frac{\int d^D x\ x_i x_j\ e^{-\frac{1}{2} \vec x\ ^t M \vec x}}{\int d^D x\ e^{-\frac{1}{2} \vec x\ ^t M \vec x}}
\end{eqnarray}
Thus,  we can write as
\begin{eqnarray}
    &&(L^+)_{ij}=\lim_{\delta \rightarrow 0} \frac{\int d^D x (x_{i} x_j) e^{-\frac{1}{2}\sum_{rt} x_r ( B G B^t+\delta I)_{rt} x_t}}{\int d^D x  e^{-\frac{1}{2}\sum_{rt} x_r (B G B^t +\delta I)_{rt} x_t}}\nonumber \\
    &=&\lim_{n\rightarrow 0}\lim_{\delta \rightarrow 0} \int d^D x d^{Dn} y\ (x_{i} x_j)\cdot\nonumber \\
    &&\hspace{.5cm}\cdot e^{-\frac{1}{2}\sum_{rt} x_r (B G B^t +\delta I)_{rt} x_t-\frac{1}{2}\sum_{m=1}^{n-1}\sum_{rt} y_r^m (B G B^t+\delta I)_{rt}y_t^m} \nonumber
\end{eqnarray}
where in the second line we used the replica trick formula \cite{Bartolucci2021}:
\begin{eqnarray}
    \frac{A}{B}=\lim_{n\rightarrow 0}A B^{n-1} .
\end{eqnarray}
We can now write
\begin{eqnarray}
    \sum_\gamma \sum_{rt} x_r B_{r\gamma} g_\gamma B_{t \gamma } x_t&=&\sum_{\gamma} g_{\gamma} \sum_{rt } x_r B_{r\gamma}x_t B_{t\gamma}\nonumber \\
    &\equiv& \sum_{\gamma} g_\gamma J_\gamma.
\end{eqnarray}
We can perform a similar trick for the other term, and we have 
\begin{eqnarray}
    \mathcal J_{\gamma}=\sum_{rt } x_r B_{r\gamma}x_t B_{t\gamma}+\sum_{m=1}^{n-1}\sum_{rt } y_r^m B_{r\gamma}y^m_t B_{t\gamma}
\end{eqnarray}
Note that the sum over $\gamma$ is the sum over all the pairs $(i,j)$ of the graph.  We have
\begin{equation}
    g_\gamma=(\Sigma_G+\Delta_G {s}_{\gamma(1)} {s}_{\gamma(2))})
\end{equation}
with $\Sigma_G=(\frac{1}{2}\text{diag}(G_f^\gamma+G_p^\gamma)$ and $\Delta_G=\frac{1}{2}\text{diag}(G_f^\gamma-G_p^\gamma)$.
Thus, we can write the integral as
\begin{eqnarray}
    \sum_{\gamma} g_\gamma \mathcal J_\gamma&=&\sum_{\gamma} \mathcal J_{\gamma}(\Sigma_G^\gamma+\Delta_G^\gamma {s}_{\gamma(1)} {s}_{\gamma(2)})\\
    &=&\sum_{\gamma} \underbrace{\Sigma_G^\gamma \mathcal J_{\gamma}}_{\mathcal Q_{ij}}\nonumber +\sum_{\gamma } \underbrace{\Delta_G^\gamma \mathcal J_{\gamma}}_{\tilde {\mathcal J}_{ij}} s_{\gamma(1)} s_{\gamma(2)}\nonumber
\end{eqnarray}
Let us now separate the second and third terms from the integral.
Since we are interested in averaging over the coupling distribution, we have the following distribution
\begin{eqnarray}
    p(s)=\frac{1}{Z_0}e^{-\beta H_0}
\end{eqnarray}
where 
\begin{equation}
H = -\sum_{i,j} J_{ij} s_i s_j - \sum_i h_i s_i
\label{eqn:IsingHamiltonian}
\end{equation}
and $h_i$ is an external field. 
We then have
\begin{eqnarray}
    &&\langle L^+_{ij}\rangle_\beta \nonumber\\ &&=\lim_{n\rightarrow 0}\lim_{\delta\rightarrow 0} \int d^D x d^{Dn} y\ x_i x_j e^{-\frac{\delta}{2}\sum_{r} x_r^2}\nonumber \\
    &&\hspace{1cm} \cdot e^{-\frac{\delta}{2}\sum_{m=1}^{n-1}\sum_{r} (y_r^m)^2-\frac{1}{2}\sum_{(i,j)} \mathcal Q_{ij}}   \nonumber \\
    &&\hspace{1cm}\cdot \langle  e^{-\frac{1}{2}\sum_{i\neq j} \tilde{\mathcal J_{i j}} s_i s_j}\rangle_\beta\label{eq:meanfield}
\end{eqnarray}
We see that the term being averaged can be interpreted as a perturbation
\begin{eqnarray}
\delta H=   -\frac{1}{2}\frac{1}{\beta}\sum_{ij}  \tilde{\mathcal J}_{ij} s_i s_j
\end{eqnarray} 
We can then write $H_1=H_0+\delta H$. As a result, we have the following expression
\begin{eqnarray}
   &&\langle L^+_{ij}\rangle_\beta =\nonumber \\
   &&\hspace{0.1cm}\lim_{n\rightarrow 0}\lim_{\delta\rightarrow 0} \int d^D x d^{Dn} y \ x_i x_j \nonumber \\
   &&\hspace{0.5cm}\cdot e^{-\frac{\delta}{2}\sum_{r} x_r^2 -\frac{\delta}{2}\sum_{m=1}^{n-1}\sum_{r} (y_r^m)^2-\frac{1}{2}\sum_{(i,j)} \mathcal Q_{ij}}\nonumber \\
   &&\hspace{0.5cm}\cdot e^{-\beta (F_1(x,y)-F_0)}\nonumber 
\end{eqnarray}
where $Z_1=\sum_{\{s_i\}} e^{-\beta (H_0+\delta H)}=e^{-\beta F_1}$ and similarly $Z_0=\sum_{\{s_i\}} e^{-\beta H_0}=e^{-\beta F_0}$.
We will now assume a mean-field theory and a Saddle-Point approximation for both $Z_0$ and $Z_1$, where we assume that $ \tilde{\mathcal J}_{ij}$ is constant.

Calculating this integral is now complicated for arbitrary interactions. We will now make an assumption about the interacting model.
The Hamiltonian for the general Ising model is given by equation \ref{eqn:IsingHamiltonian}, where $ s_i = \pm 1 $ are the spin variables, $ J_{ij} $ are the interaction strengths between spins, and $ h_i $ are the external fields acting on the spins. The key problem is that this model cannot be solved for arbitrary coupling constants $J_{ij}$ and field $h_i$. We will then consider an approximation of this model given by a mean-field approach.

In the Curie-Wei\ss\  model, the interactions are uniform and mean-field-like, meaning each spin interacts equally with every other spin. The Hamiltonian for the Curie-Wei\ss\  model is:
\begin{eqnarray}
H_{mft} &=& -\frac{J_{eff}}{2N} \left( \sum_{i=1}^N s_i \right)^2 - h_\text{eff} \sum_{i=1}^N s_i \nonumber \\
&=& -N\left(\frac{J_{eff}}{2} m^2 + h_\text{eff} m\right)\label{eq:mft00}
\end{eqnarray}
where $m=\sum_{i=1}^N s_i/N$. 

To map the general Ising model to the Curie-Wei\ss\  model, we derive effective parameters $ J $ and $ h $ that correspond to the average behavior of the spins.
 In the mean-field approximation, we replace the interaction term with an effective field that each spin experiences due to the average magnetization $ m = \langle s_i \rangle $ where the deviation between each spin and the average value is a small parameter
\begin{equation}
s_i s_j \approx \langle s_i \rangle s_j + s_i \langle s_j \rangle - \langle s_i \rangle \langle s_j \rangle\label{eq:mft0}
\end{equation}

The average interaction strength in the Curie-Wei\ss\  model is:
\begin{equation}
J_{\text{eff}} = \frac{1}{N} \sum_{i \neq j} (J_{ij}-\frac{1}{\beta}\tilde{\mathcal J}_{ij})
\end{equation}
Here, the factor of $\frac{1}{N}$ accounts for the mean-field nature of the Curie-Wei\ss\  model, where each spin interacts equally with all other spins.

Thus, the first approximation is that we replace Eqn. (\ref{eq:meanfield}) with
\begin{eqnarray}
    \langle \cdot \rangle_\beta\approx \langle \rangle_{mft},
\end{eqnarray}
where
\begin{eqnarray}
    \langle\ \cdot\ \rangle_{mft}= \sum_{\{s_i\}}\Big(\ \cdot\ \Big) \frac{e^{-\beta H_{mft}}}{Z_{mft}},
\end{eqnarray}
where $Z_{mft}=\sum_{\{s_i\}}e^{-\beta H_{mft}}$.
Thus, we can use the mean field theory approximation to infer the properties of the resistance, implementing 
\begin{eqnarray}
    &&\langle  e^{\frac{1}{2}\sum_{i\neq j} \tilde{\mathcal J}_{i j} s_i{s}_j}\rangle_\beta\nonumber \\
    &&\hspace{1cm}\approx\langle  e^{\frac{1}{2}\sum_{i\neq j} \tilde{\mathcal J}_{i j} {s}_i\cdot {s}_j}\rangle_{mft}\\
    &&\hspace{1cm}\approx e^{\frac{1}{2}\sum_{i\neq j}  \tilde{\mathcal J}_{i j} \langle {s}_i\cdot {s}_j\rangle_{mft}}
\end{eqnarray}
and using eqn. (\ref{eq:mft0}), we have
\begin{eqnarray}
    \langle s_i s_j\rangle_{mft}&=&\langle s_i\rangle_{mft} \langle s_j\rangle_{mft}\nonumber \\
    &=&m^2
\end{eqnarray}
where $m$ is determined via the mean-field equation:
\begin{eqnarray}
    m=\tanh\big(\beta(J_{eff} m+h_\text{eff})\big)\label{eq:mftfin}
\end{eqnarray}
and is interpreted as the mean magnetization arising from the  Curie-Wei\ss\ model of eqn. (\ref{eq:mft00}) applied to the average of eqn. (\ref{eq:meanfield}.

Under these conditions it is easy to see then that we can unpack all the limits, we can write%
\begin{eqnarray}
    \langle L^+_{ij}\rangle &\approx ( B\tilde{G}(m,\phi) B^t)^+\label{eq:mft}
\end{eqnarray}


This is the analytical formula we will use in the random graph models where $m$ satisfies eqn. (\ref{eq:mftfin}).

\begin{figure*}
    \centering
    \includegraphics[width=\linewidth]{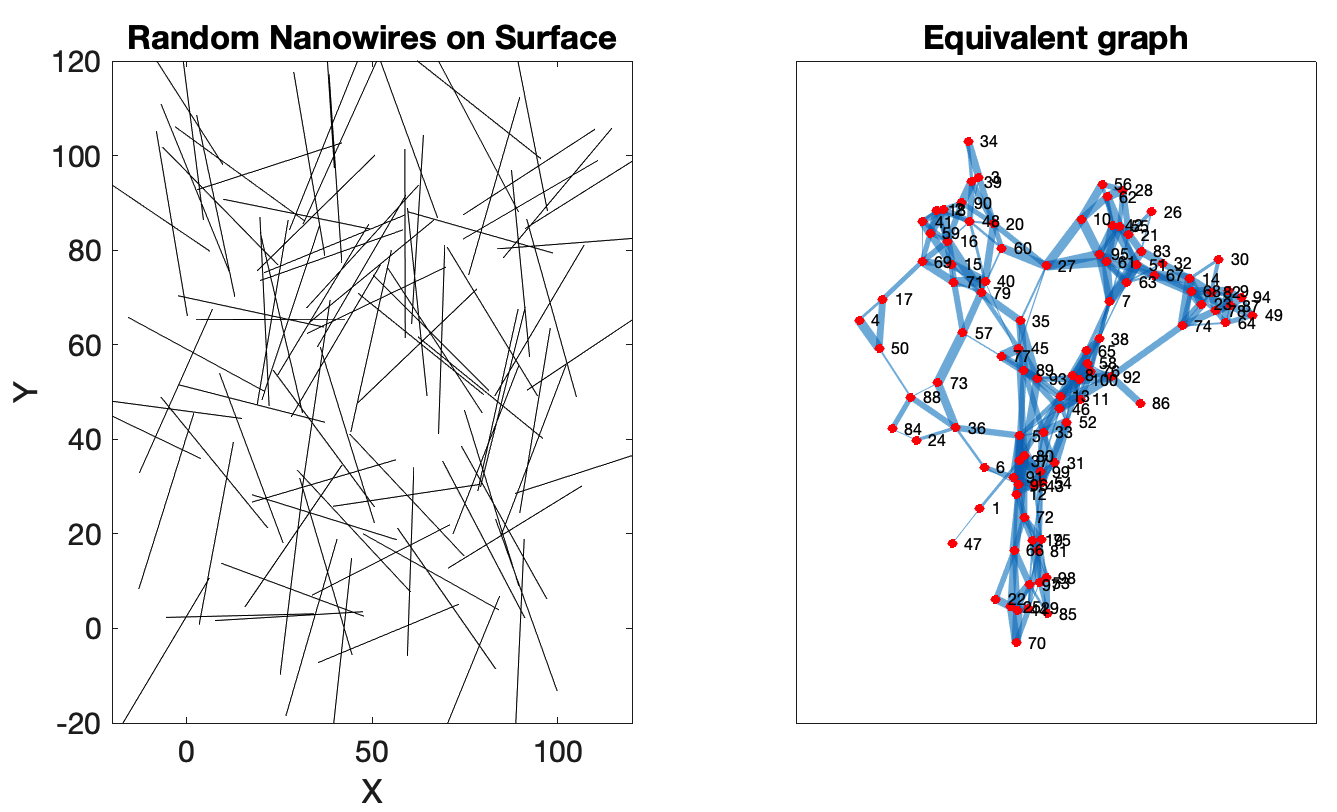}
    \caption{Example of the random nanowire network and the equivalent graph constructed using the method of Figure \ref{fig:junctmap}. Each nanowire overlap is converted to a resistive junction whose width depends on $\cos(\phi_{ij})$.}
    \label{fig:randomnanowire}
\end{figure*}

\section{Random graph models}
\subsection{Thermal average}
The Monte Carlo Metropolis-Hastings algorithm is used to generate samples from a target probability distribution $\pi(x)$. We just briefly recall the method \cite{Metropolis1953,Landau2021}. Starting from an initial state $x_0$, a candidate state $x'$ is proposed from a distribution $q(x'|x)$, the $x^\prime$ states accessible from $x_0$. The acceptance probability is computed as
\begin{eqnarray}    
\alpha = \min \left(1, \frac{\pi(x') q(x|x')}{\pi(x) q(x'|x)} \right).
\end{eqnarray}
The candidate state $x'$ is accepted with probability $\alpha$; otherwise, the current state is retained. This process is repeated, creating a Markov chain that converges to the equilibrium distribution $\pi(x)$. The algorithm ensures detailed balance, $\pi(x) q(x'|x) \alpha = \pi(x') q(x|x')$, which guarantees that the chain's stationary distribution is $\pi(x)$. With sufficient iterations, it is known the samples represent the equilibrium distribution, making this method effective for sampling from complex distributions.

In the case of the Ising system, the accessible states $x^\prime$ correspond to states that have undergone a physical process, typically single spin flips. The acceptance probability then depends on a function, e.g., the Boltzmann distribution. By progressively lowering the temperature, lower energy spin configurations are identified until the low-temperature equilibrium distribution corresponds to the ground state distribution $\pi(x)$.

\subsection{Random graphs}\label{sec:er}
\begin{figure}
    \centering
    \includegraphics[width=\linewidth]{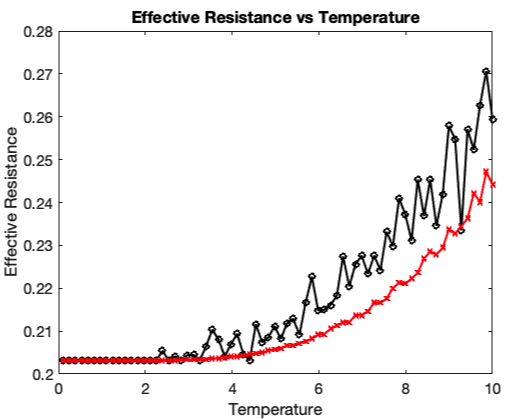}
    \caption{Example of mean effective resistance as a function of temperature for the Erd\'os-R\`enyi random graphs $G(n,m)$, with $n=50$, $m=280$ from Sec. \ref{sec:er}, with the resistance parameters $R_f=1$, $R_a=2$. The black line is calculated using a Metropolis-Hasting algorithm and averaged over the last 10 points of the simulation, with $500$ Monte Carlo steps. The red curve is obtained from eqn. (\ref{eq:mft}) via the mean-field theory approach.}
    \label{fig:randomgraph}
\end{figure}
In the Erdős-Rényi model $\mathcal{G}(n, m)$, a graph is generated by randomly selecting $m$ edges from the $\binom{n}{2}$ possible edges between $n$ nodes. The construction of the Erdős-Rényi random graph is well known and can be described as follows \cite{Bollobs2001}: a) Start with a set of $n$ nodes, denoted as $V = \{v_1, v_2, \ldots, v_n\}$ b)   Randomly select $m$ edges from the set of all possible edges $\binom{n}{2}$. Each possible edge $(v_i, v_j)$ is chosen with equal probability. c)   The ensemble $\mathcal{G}(n, m)$ consists of all possible graphs with $n$ nodes and $m$ edges, each occurring with equal probability:
   \begin{equation}
   \mathcal{G}(n, m) = \{\mathcal{G} = (\mathcal{V}, E) \mid |\mathcal{V}| = n, |E| = m\}.
   \end{equation}
The Erdős-Rényi random graph model has several important properties. First, the degree of a node follows a binomial distribution $ \text{B}(n-1, p) $, where $ p = \frac{2m}{n(n-1)} $ is the probability of an edge existing between any two nodes.
    For large $n$, if $m > \frac{n \ln n}{2}$, the graph almost surely contains a giant component, which is a connected subgraph that includes a positive fraction of the nodes.   If $m > \frac{n \ln n}{2}$, the graph is almost surely connected, which is useful for our context.
For these types of graphs, and assuming that $\phi_{ij}$ is uniformly random on $[0,2\pi]$, we have the mean effective resistance of Fig. \ref{fig:randomgraph} at low temperatures. We see that our mean field theory captures the behavior of the effective resistance in this case.

\begin{figure}
    \centering
    \includegraphics[width=\linewidth]{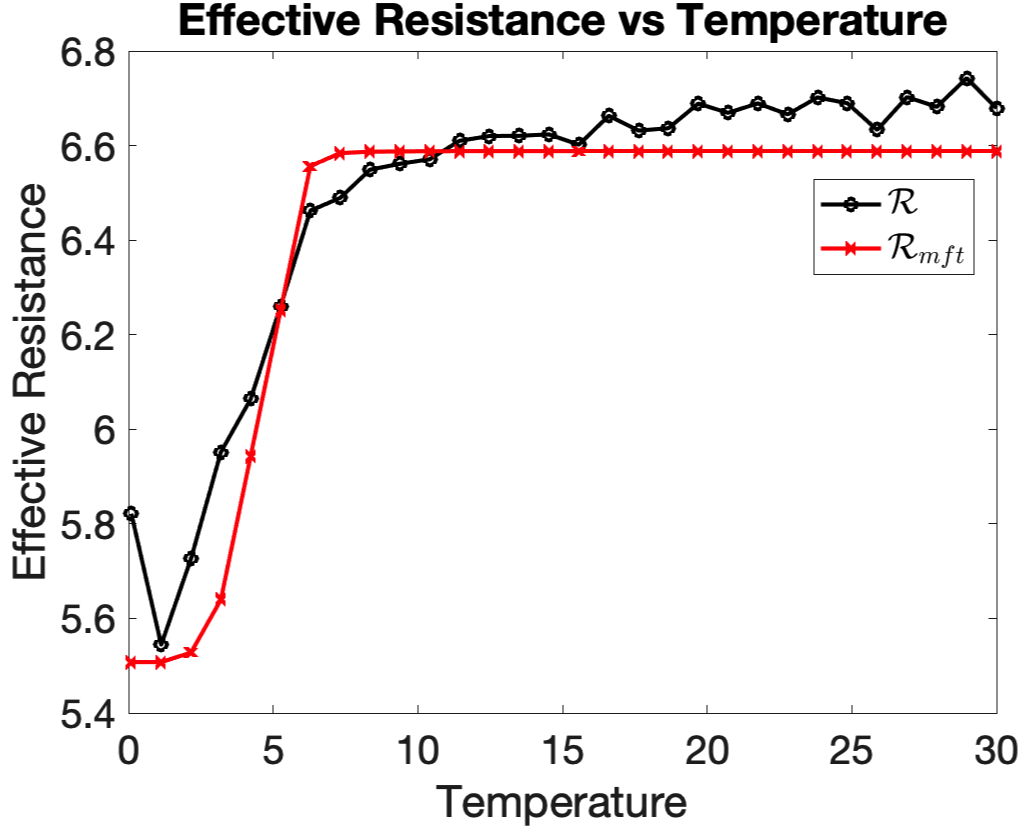}
    \caption{Effective resistance calculated with Monte Carlo sampling, for $N=100$ nanowires of length $40\mu$m on a surface of $100^2 \mu m^2$ using the generation model of Sec. \ref{sec:randomnw}. The relaxation was performed for $1000N$ Monte Carlo steps, and each point is averaged over the last $100$ points of the simulation. Again, the red curve is obtained from eqn. (\ref{eq:mft}) via the mean-field theory approach. In this calculation, we have set $J=1$, and thus the temperature is in units of the effective interaction constant between the nanowires. This method is valid if the dipolar interaction is negligible compared to the domain walls generation.}
    \label{fig:effresnw}
\end{figure}

\begin{figure}[h!]
    \centering
    \includegraphics[width=0.97\linewidth]{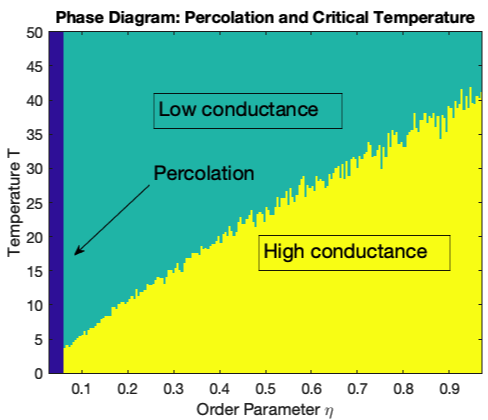}\\
    \includegraphics[width=0.97\linewidth]{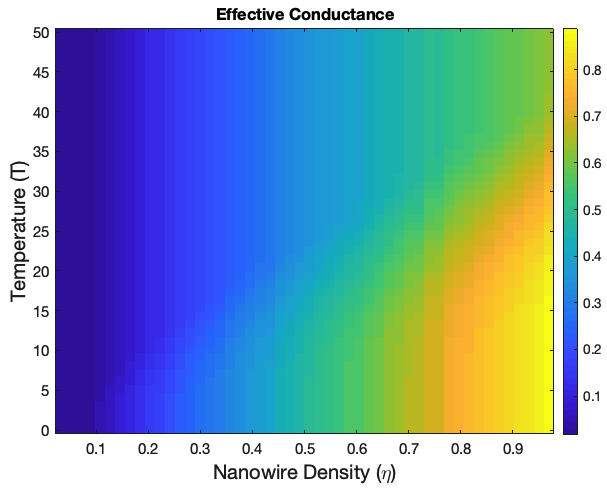}\\
    \caption{Numerically obtained phase diagram as a function of the order parameter $\eta$ and temperature $T$, with up to $600$ nw.  The resistance choices for each nanowire junction were those of eqn. (\ref{eq:wij}). In both the plots above, $\eta$ was obtained and calculated via eqn. (\ref{eq:eta}). On the horizontal axis (top and bottom diagrams), the critical line was calculated based on a randomly generated nanowire sample given $\eta$. A single instance obtained the top plot for each value of $\eta$, while the bottom plot's curves were averaged over $20$ randomly generated samples which were equilibrated using a Metropolis algorithm at the fixed temperature.  The color scheme indicates the percolation and resistance state of the nanowire network: blue for non-percolating, light green for percolating with low conductance (high resistance), and yellow for percolating with high conductance (low resistance). These results are obtained in the absence of external magnetic fields. In the top figure, we estimate the transition temperature as a function of the $J_{eff}$.}
    \label{fig:PhaseDiagram}
\end{figure}

\subsection{Random nanowires model}\label{sec:randomnw}

Let us now discuss a more realistic model for the graphs, based on the deposition of nanowires on a sample. The algorithm that we introduce below is a Monte Carlo in which at each time step, a wire of a certain length is deposited on a surface, with the center and orientation chosen at random \cite{Zhu2021,Milano2022a}. 
The algorithm for generating the random graph from the random nanowire network can be described in the following steps:

\begin{enumerate}
    \item  Randomly place nanowires on a surface. Each nanowire is represented by a line segment of fixed length $l$ and a random orientation $\theta$, uniformly distributed between $0$ and $2\pi$.

   \item For each pair of nanowires, determine if they intersect. The intersection of two line segments is checked using the parametric form of the line equations.

   \item If two nanowires intersect, calculate the relative angle $\phi$ between them. This angle is used to determine the weight of the edge in the graph.

   \item Finally, we construct a graph where each nanowire corresponds to a node. Add an edge between two nodes if the corresponding nanowires intersect. The weight of the edge is determined by the relative angle between the intersecting nanowires.
\end{enumerate}
Let $\mathbf{N} = \{N_1, N_2, \ldots, N_n\}$ represent the set of $n$ nanowires, where each nanowire $N_i$ is defined by its endpoints $(x_i^1, y_i^1)$ and $(x_i^2, y_i^2)$.

The intersection condition between two nanowires $N_i$ and $N_j$ can be expressed using the parametric line equations:
\begin{align}
(x_i, y_i) &= (x_i^1, y_i^1) + t_i \left((x_i^2, y_i^2) - (x_i^1, y_i^1)\right), \quad t_i \in [0, 1], \\
(x_j, y_j) &= (x_j^1, y_j^1) + t_j \left((x_j^2, y_j^2) - (x_j^1, y_j^1)\right), \quad t_j \in [0, 1].
\end{align}
The intersection point $(x_0, y_0)$ must satisfy the equation:
\begin{eqnarray}
&&(x_i^1, y_i^1) + t_i \left((x_i^2, y_i^2) - (x_i^1, y_i^1)\right) \nonumber \\
&&\hspace{0.5cm}= (x_j^1, y_j^1) + t_j \left((x_j^2, y_j^2) - (x_j^1, y_j^1)\right).
\end{eqnarray}

If the intersection exists, we calculate the relative angle $\phi_{ij}$ between the nanowires $N_i$ and $N_j$ as:
\begin{equation}
\phi_{ij} = \arccos \left( \frac{\mathbf{d}_i \cdot \mathbf{d}_j}{|\mathbf{d}_i| |\mathbf{d}_j|} \right),
\end{equation}
where $\mathbf{d}_i = (x_i^2 - x_i^1, y_i^2 - y_i^1)$ and $\mathbf{d}_j = (x_j^2 - x_j^1, y_j^2 - y_j^1)$ are the direction vectors of the nanowires.

We now construct a graph $\mathcal{G} = (\mathcal{V}, E, W)$ where: $\mathcal{V}$ is the set of nodes, each corresponding to a nanowire, $E$ is the set of edges, with an edge $e_{ij} \in E$ if nanowires $N_i$ and $N_j$ intersect. $W$ is the set of edge weights, where the weight $w_{ij}(s_i,s_j)$ of edge $e_{ij}$ is dependent on the cosine of the angle $\phi_{ij}$:
\begin{equation}
w_{ij} = \mathcal A_{ij}+\mathcal B_{ij} \cos(\phi_{ij}) \label{eq:wij}
\end{equation}
This is the formula we consider in our numerical simulations, with $\mathcal A\approx 20$ Ohms, and $|B/A|\approx 0.5$, compatible with the value obtained from the 
tunneling magnetoresistance effect. Specifically, $R_{++}=a-b \cos(\phi_{ij})$, while $R_{+-}=a+b \cos(\phi_{ij})$, e.g. only $\mathcal B$ is spin dependent.

\section{Analysis}

An example of a random nanowire network generated with 100 nanowires is shown in Figure \ref{fig:randomnanowire}, with the equivalent interaction graph constructed on the right-hand side. For this example, we selected a graph that is fully connected but depending on the density of nanowires $\eta$, which we defined as the number of nanowires multiplied by the length of the nanowire squared and divided by the area of the surface over which they are deposited, an equivalent graph can have various disconnected components. In Figure \ref{fig:effresnw}, we show the mean effective resistance calculated numerically (black line) as a function of temperature, assuming that the single junction's parameter has $\mathcal A_{ij}=20$ Ohms, and $\mathcal B_{ij}=10$ Ohms. We see that the effective network resistance varies in the range $\approx 5.6-7$ Ohms in this case, with $100$ nanowires for that particular simulation. We can see a good match with the mean-field theory developed earlier, which justifies the use use of the network-dependent critical temperature $T_{c}=J_{eff}^{-1}$ as our probe. In general, this critical temperature will depend on the density of nanowires, as this will affect the mean network connectivity.

Here, the interaction is assumed to be local at the junction, and captures the effective generation of walls at the junction which causes uniform magnetization 'spin' flips. If dipolar interaction is present and strong, then the mean-field method we use will not be capturing the underlying physics, as the effective system will be frustrated, e.g., the interaction between the islands can be ferromagnetic or antiferromagnetic depending on the relative angle.

To investigate the critical resistance in nanowire networks, we constructed a phase diagram as a function of the order parameter $\eta$ and temperature $T$. This phase diagram delineates regions of percolation and resistance properties within the network.

The order parameter $\eta$ is defined as the ratio of the total length of the nanowires to the area of the surface:
\begin{equation}
\eta = \frac{N \cdot l^2}{L^2}\label{eq:eta}
\end{equation}
where $N$ is the number of nanowires, $l$ is the length of each nanowire, and $L$ is the size of the surface. To study the percolation threshold and resistance properties, we varied the length of the nanowires according to the equation $l = c_0 \sqrt{\frac{\eta \cdot L^2}{N}}$
where $c_0$ is a proportionality constant and $N = 600$ nanowires for our specific plot. This choice ensures that the nanowires' length appropriately scales with $\eta$, keeping the surface area and number of nanowires constant. 

In the mean field theory approximation, the critical temperature $T_c$ for the network is obtained from the effective coupling $J_{\text{eff}}$. The effective coupling is calculated using the adjacency matrix of the graph representing the nanowire network:

\begin{equation}
J_{\text{eff}} = \frac{\sum_{i,j} \left| A_{ij} \right|}{N}
\end{equation}
The critical temperature is then given by
\begin{equation}
T_c=\kappa_B J_{\text{eff}}
\end{equation}
This relationship is derived from the mean-field theory, which approximates the behavior of the network's resistance above the percolation threshold. 
One can interpret this temperature physically similarly to the temperature introduced in (thermal) artificial spin ice \cite{nisoliefft}, as a parameter controlling the probability of the magnetization to follow the ``dipolar"-induced ground state. Now, note that the parameter $\eta$ controls the density of nanowires per units of area. If either the number of NW or the length of NW increases, one should expect more overlaps and thus a higher average degree. We should expect a critical temperature that depends directly on the density of nanowires.  With this in mind, we can now study the phase diagram of this model.

To generate the phase diagram, we employed a Monte Carlo simulation. For each value of $\eta$, we created a random nanowire network and computed its adjacency matrix, and $J_{eff}$. The percolation status of the network was determined by calculating the giant cluster and testing whether the connected component was at least half of the number of nanowires. The critical temperature was calculated as described above.

The phase diagram in Figure \ref{fig:PhaseDiagram}(a) reveals three distinct regions of behavior for up to $600$ nanowires (set so that this corresponds to $\eta=1$): a) At low $\eta$, the network is predominantly not percolating (blue region), indicating that the density of nanowires is insufficient for forming a spanning cluster. As $\eta$ increases, the network percolates (transition to green and yellow regions). The critical temperature $T_c$ increases with increasing $\eta$, indicating that denser networks require higher temperatures to transition into a low-resistance state.  Low-resistance percolating networks are shown as the green region. The boundary between light and dark green regions marks the critical temperature $T_c$ for different $\eta$ values. c) Low resistance percolating networks are shown as the yellow region. 

In Figure \ref{fig:PhaseDiagram} (b) we plot instead the mean effective conductance as a function of $\eta$ and $T$. Each point is averaged over $20$ configurations, which were then thermalized via a Metropolis Monte Carlo algorithm. We see that the rough behavior of the conductance matches the approximate phase diagram obtained via the mean-field in Fig. \ref{fig:PhaseDiagram}(a). The percolation threshold is not directly observable from the conductance as it is hidden in the blue region at low nanowire density.
 
This phase diagram provides a comprehensive view of the percolation and resistance behavior in the nanowire networks, clarifying the relationship between network density, temperature, and resistance properties. It is interesting to note the similarity of the phase diagram of Figure \ref{fig:PhaseDiagram} with the one developed in \cite{Caravelli2023} for silver nanowires. In that case, the controlling parameter was not the temperature but the effective voltage applied to the network, inducing the formation of filaments at each silver junction. However, the percolation threshold was already discussed there, as this is of a geometrical nature. The mechanism for the low and high resistance states was different there also (and dynamic in nature), while in our plot these are equilibrium configurations. Nonetheless, it is interesting to note the similarity in the structure of the phase diagram, with a critical line separating high and low resistance states.

\section{Conclusions}

In this study, we have investigated the effective resistance in random networks of magnetic nanowires, focusing on the role of magnetic tunnel junctions (MTJs) formed at the intersections of nanowires. Our analysis has revealed several key insights into the interplay between magnetic and electrical properties in these complex systems. We have demonstrated that the resistance of a junction in a magnetic nanowire network is inherently dependent on the relative magnetization states of the intersecting nanowires. This dependence is captured by a model that considers both ferromagnetic and antiferromagnetic alignments, leading to a resistance formula that effectively encapsulates the effects of magnetization states on the network's overall electrical behavior.
Specifically, we used the Moore-Penrose pseudoinverse of the Laplacian matrix to compute the effective resistance of the network. This approach allows us to incorporate the resistive properties of individual junctions and provides a comprehensive framework for understanding how the network responds to external electrical stimuli. Our results highlight the critical role of magnetic interactions in determining the effective resistance and, by extension, the performance of the network in practical applications.
Our findings underscore the importance of considering both the magnetic and electrical properties of nanowire networks in the design and optimization of amorphous resistive devices. As we have shown, the intricate interplay between these properties can significantly influence the overall behavior of the network. As a result, we have shown that this type of phase change magnetic materials is strongly affected by the temperature. Future work should focus on exploring the effects of varying the geometrical and magnetic configurations of the nanowires, incorporating magnetic domains within nanowires, as well as incorporating more complex magnetic interactions and disorder into the model. Such studies will provide deeper insights into the behavior of magnetic nanowire networks and aid in the development of advanced materials.
In conclusion, our study provides a robust theoretical framework for analyzing the resistive behavior of magnetic nanowire networks, shedding light on the fundamental mechanisms governing their electrical properties.


\begin{acknowledgments}

Our work was conducted under the auspices of the National Nuclear Security Administration of the United States Department of Energy at Los Alamos National Laboratory (LANL) under Contract No. DE-AC52-06NA25396.  F. Caravelli was also financed via DOE LDRD grant 20240245ER. This material is based upon work supported by the National Science Foundation under Grant No. 2205796 (E.I.)

\end{acknowledgments}

\bibliography{bibtex}

\clearpage
\appendix

\end{document}